\PassOptionsToPackage{english}{babel}
\documentclass[reprint,aps,prl,twocolumn,showpacs,amssymb,amsfonts, superscriptaddress,longbibliography]{revtex4-1}  
\usepackage[english,ngerman]{babel}
\usepackage{graphicx}  
\usepackage{dcolumn}   
\usepackage{bm}        
\usepackage{amssymb}   
\usepackage{amsmath}
\usepackage{mathtools}
\usepackage[utf8]{inputenc}
 
\usepackage{changes}

\definecolor{myurlcolor}{rgb}{0,0,0.7}
\definecolor{myrefcolor}{rgb}{0.8,0,0}
\usepackage{hyperref}
\hypersetup{colorlinks, linkcolor=myrefcolor,
citecolor=myurlcolor, urlcolor=myurlcolor}

\usepackage{cleveref}
\crefrangeformat{equation}{#3(#1 #4--#5 #2)#6}

\usepackage{comment}

\newcommand{\ignore}[1]{}

\begin{document}
\selectlanguage{english}

\title{Bounding the set of classical correlations of a many--body system}

\author{Matteo Fadel}\email{matteo.fadel@unibas.ch} \affiliation{Department of Physics, University of Basel, Klingelbergstrasse 82, 4056 Basel, Switzerland} 

\author{Jordi Tura} \email{jordi.tura@mpq.mpg.de} \affiliation{ICFO - Institut de Ciencies Fotoniques, The Barcelona Institute of Science and Technology, 08860 Castelldefels (Barcelona), Spain}
\affiliation{Max-Planck-Institut f\"ur Quantenoptik, Hans-Kopfermann-Stra{\ss}e 1, 85748 Garching, Germany}

\date{\today}

\begin{abstract}

We present a method to certify the presence of Bell correlations in experimentally observed statistics, and to obtain new Bell inequalities. Our approach is based on relaxing the conditions defining the set of correlations obeying a local hidden variable model, yielding a convergent hierarchy of semidefinite programs (SdP's). Because the size of these SdP's is independent of the number of parties involved, this technique allows to characterize correlations in many--body systems. As an example, we illustrate our method with the experimental data presented in [\href{https://doi.org/10.1126/science.aad8665}{Science {\bf 352}, 441 (2016)}].


\end{abstract}

\maketitle

\paragraph{Introduction.}
-- Local measurements on quantum systems can display correlations that can not be explained by any local hidden variable model (LHVM) \cite{Bell64} or, in other words, that can not be reproduced by local deterministic strategies (LDS), even if assisted by shared randomness \cite{Fine82}. Bell inequalities bound the space of LHVM or ``classical'' correlations, and correlations that violate a Bell inequality are termed nonlocal. Beside their fundamental interest, nonlocal correlations are a resource that enables novel quantum information processing tasks \cite{ReviewNonlocality}.

From a geometrical point of view, LHVM correlations form a polytope, \textit{i.e.}\ a bounded convex set that can be described as the convex hull of a finite number of vertices, or equivalently as the intersection of a finite number of half-spaces. The vertices of the LHVM polytope correspond to LDS, while the half-spaces in which it is contained are defined by Bell inequalities. For this reason, finding all Bell inequalities gives a necessary and sufficient condition for deciding membership in the LHVM set. However, results in computer science indicate that this search is an extremely demanding problem \cite{Babai1991}, which is NP-complete even in the bipartite case \cite{Avis04}. Therefore, a complete list of Bell inequalities exists only for the simplest scenarios; \textit{e.g.}\ only up to $3$ parties \cite{faacet, CHSH, PironioCHSH, Pitowsky01, Sliwa03}.

To characterize correlations in scenarios with a large number of parties, one necessarily has to relax the condition of membership in the LHVM set. This can be done by projecting the LHVM polytope onto the space of observables of a particular form, \textit{e.g.}\ permutationally invariant \cite{BancalSymmetricIneq}, with low-order correlators \cite{TuraSCIENCE14, TuraAnnPhys}, or translationally invariant \cite{TIpaper, EnergyNonlocality}. Finding Bell inequalities in these particular spaces has allowed the detection of Bell correlations in a Bose-Einstein condensate of $480$ $^{87}\mathrm{Rb}$ atoms \cite{SchmiedSCIENCE16}. 
However, even in these low dimensional spaces, the complexity of the commonly adopted method (going from the vertices description of the polytope, to the dual half-spaces description) \cite{Chazelle93} still prohibits one to obtain all Bell inequalities for many-body scenarios, leaving undiscovered potentially useful inequalities.

In this work, we present a technique to approximate the set of symmetric LHVM correlations from the outside. This technique is based on a hierarchy of semidefinite programs (SdP's), aproximating convex hulls of semialgebraic sets \cite{Gouveia10, GouveiaBOOK12, LasserreBook,LasserreBook2}, and it can be seen as checking all Bell inequalities of a specific form with a single test. Contrary to other existing SdP's hierarchies \cite{Baccari16,NPA}, in our work the size of the SdP's are independent of the number of parties, and the hierarchy shows convergence already after few levels. In summary, our method provides an efficient sufficient condition for a set of correlations to be nonlocal, and it naturally provides a Bell inequality that they violate.

\paragraph{Preliminaries.}
-- We consider a Bell scenario in which each of $N$ observers (indexed by $i = 1\ldots N$) performs on their share of the system one out of $d$ possible local measurements $\mathcal{M}_{j}^{(i)}$, labeled by $j=0\ldots d-1$. For simplicity, we assume that every measurement is dichotomic, giving as outcome $+1$ or $-1$, keeping the generalization to an arbitrary number of outcomes for later. The correlations that can be observed are represented by the $k$-body correlators $\langle \mathcal{M}_{j_1}^{(i_1)}...\mathcal{M}_{j_k}^{(i_k)} \rangle$, with $k=1,...,N$. In the spirit of \cite{BancalSymmetricIneq, TuraSCIENCE14, TuraAnnPhys}, we focus on permutationally invariant (PI) $k$-body correlators, defined as

\begin{equation}
\mathcal{S}_{j_1 ... j_k} = \sum_{\substack{i_1,...,i_k=1 \\ \text{all $i$'s different} }}^N \langle \mathcal{M}_{j_1}^{(i_1)}...\mathcal{M}_{j_k}^{(i_k)} \rangle \;.
\label{symcorr}
\end{equation}
If the statistics observed through Eq.~(\ref{symcorr}) satisfy a LHVM, they belongs to the so-called (symmetrized) LHVM polytope \cite{BancalSymmetricIneq}, denoted $\mathbb{P}^\text{S}$. Its vertices correspond to LDS, satisfying
\begin{equation}\label{local}
\langle \mathcal{M}_{j_1}^{(i_1)}...\mathcal{M}_{j_k}^{(i_k)} \rangle = \langle \mathcal{M}_{j_1}^{(i_1)}\rangle \cdots \langle\mathcal{M}_{j_k}^{(i_k)} \rangle \qquad\text{(local)} \;,
\end{equation}
\begin{equation}\label{determ}
\langle \mathcal{M}_{j}^{(i)}\rangle = \pm 1 \quad\forall\; i,j \qquad\text{(deterministic)} \;.
\end{equation}

As there are $m=2^d$ possible LDS per party, Eq.~(\ref{determ}) gives rise to an exponential number of vertices, $2^{dN}$. However, the PI condition reduces them to at most ${N + m-1} \choose {m -1}$, a polynomial number in $N$, because only the amount of parties following the same LDS is relevant \cite{TuraSCIENCE14,TuraAnnPhys}.
For this reason, it is natural to introduce $m$ variables $\vec{x} = (x_1, \ldots, x_{m})$, where $x_i$ counts how many parties follow the $i$-th LDS. Note that the $x_i$ satisfy
\begin{equation}
\sum_{i=1}^{m} x_i = N, \qquad x_i \in \mathbb{Z}_{\geq 0}.
\label{parameters}
\end{equation}
Using this parameterization, Eq.~(\ref{symcorr}) can be written as a polynomial of degree $k$ in $m$ variables with real coefficients, \textit{i.e.}\ $\mathcal{S}_{j_1 ... j_k} \in \mathbb{R}[\vec{x}]_k$ (see \cite{TuraSCIENCE14} and the example).
Denoting with $\vec{\cal S}$ the vector of all such correlations, we express $\mathbb{P}^{\text{S}}$ as the convex hull (CH) of $\vec{\cal S}$ evaluated on the parameter region defined by Eq.~(\ref{parameters}):
\begin{equation}
\mathbb{P}^{\text{S}} = \text{CH}\left\lbrace \vec{\cal S}(\vec{x}) \;\;\text{s.t.}\;\; \sum_i x_i = N,\  x_i \in {\mathbb{Z}_{\geq 0}}\right\rbrace.
\label{eq:PsCHZ}
\end{equation}

Dedicated algorithms \cite{PORTA, CDD} exist to compute the dual description of $\mathbb{P}^{\text{S}}$, thus obtaining a minimal set of PI Bell inequalities. These inequalities are of the form
\begin{equation}
\sum_k \sum_{j_1\leq\ldots \leq j_k} \alpha_{j_1 \ldots j_k} {\cal S}_{j_1\ldots j_k} + \beta_C \geq 0,
\label{eq:BI}
\end{equation}
where $\alpha_{j_1 \ldots j_k} \in \mathbb{R}$, and $\beta_C\in \mathbb{R}$ is the so-called classical bound. Unfortunately, since the dimension of $\mathbb{P}^{\text{S}}$ scales as ${d + N \choose d} -1$, one in practice can not obtain a full set of BI for $N > 5$ \cite{BancalSymmetricIneq}. However, it has recently been shown both theoretically \cite{TuraSCIENCE14, TuraAnnPhys}, and experimentally \cite{SchmiedSCIENCE16}\ignore{EngelsenPRL}, that a small subset of the correlators in $\vec{\cal S}$ (namely, one- and two-body PI correlators) suffices to detect Bell correlations for arbitrarily large $N$. Therefore, we limit the number of components of $\vec{\cal S}$ to contain only up to $K$-body correlators, effectively projecting $\mathbb{P}^{\text{S}}$ to a polytope $\mathbb{P}^{\text{S}}_K$ living in a subspace of dimension ${d + K \choose d} -1$, independent of $N$ , whose vector of coordinates we denote $\vec{\cal S}_K$.
Still, in the case $K=2$ finding all PI BI only works for $N \lesssim 40$ in less than a month runtime. Hence, to study the large $N$ regime one has to (i) infer classes of BI and generalize them to arbitrary $N$ and (ii) derive a proof of their $\beta_C$ for each class. Nevertheless, as more BI appear as $N$ increases, this procedure may leave potentially more useful classes unnoticed if they do not show up for sufficiently small $N$.

We propose here a method to approximate $\mathbb{P}^{\text{S}}_K$ from the outside, which overcomes the above limitations. Our technique is based on two mild relaxations yielding a hierarchy of conditions satisfied by all LHVM correlations.

\paragraph{First relaxation.} -- According to Eq.~(\ref{eq:PsCHZ}), $\mathbb{P}^{\text{S}}$ is defined as the convex hull of a finite set of points, therefore not exploiting the inherent algebraic structure present in the polynomials $\vec{\cal S}_K(\vec{x})$. The first relaxation we introduce consists in dropping the condition $x_i \in \mathbb{Z}_{\geq 0}$, and consider instead $x_i \in \mathbb{R}_{\geq 0}$, which gives rise to the set
\begin{equation}
\widetilde{\mathbb{P}^{\text{S}}_K} = \text{CH}\left\lbrace\vec{\cal S}_K(\vec{x}) \;\;\text{s.t.}\;\; \sum_i x_i = N,\  x_i \in {\mathbb{R}_{\geq 0}}\right\rbrace \;.
\label{eq:PsCHR}
\end{equation}
Note that $\vec{\cal S}_K(\vec{x})$ with $\vec{x} \in \mathbb{R}^m$ interpolates the vertices of $\mathbb{P}^{\text{S}}_K$, implying $\mathbb{P}^{\text{S}}_K \subseteq \widetilde{\mathbb{P}^{\text{S}}_K}$. As a consequence, if a set of correlations lies outside $\widetilde{\mathbb{P}^{\text{S}}_K}$, it also lies outside $\mathbb{P}^{\text{S}}_K$, and therefore it is nonlocal.

Since $\vec{x}$ has $m-1$ free parameters, and $\vec{\cal S}_K$ has ${d + K \choose d}-1$ components, $\vec{\cal S}_K(\vec{x})$ can be expressed as a set of equations $f_i(\vec{S}_K)=0$, where $1\leq i \leq {d+K \choose d}-m$. The non-negativity constraints $x_j \geq 0$ can also be expressed as a set of $m$ constraints in $\vec{\cal S}_K$, by a set of inequalities $g_j(\vec{\cal S}_K) \geq 0$ (see the example). 
In what follows, we refer to the set of solutions of a system of polynomial equations $f_i(\vec{\cal S}_K) = 0$ as an algebraic set. Moreover, if an algebraic set is further restricted by polynomial non-negativity constraints $g_j(\vec{\cal S}_K) \geq 0$, as it is the case for $\widetilde{\mathbb{P}^{\text{S}}_K}$ in Eq.~(\ref{eq:PsCHR}), we shall call such a set semialgebraic.

\paragraph{Second relaxation.} -- Deciding membership in the CH of a (semi)algebraic set $\cal V$ is NP-hard  \cite{LasserreBook}. However, there exist efficient approximations for $\mathrm{CH}({\cal V})$ from the outside \cite{LasserreBook,LasserreBook2, Gouveia10, GouveiaBOOK12}. The idea behind these methods is to reduce the membership problem in $\mathrm{CH}({\cal V})$ to that of a multivariate polynomial being non-negative, which can be relaxed to determining whether such polynomial can be expressed as a sum of squares (s.o.s.) \footnote{Obviously, every s.o.s.\ polynomial is non-negative, but the converse is false in general. The textbook counter-example is the Motzkin polynomial, $x^4y^2 + x^2y^4 - 3x^2y^2 + 1$, which is non-negative on $\mathbb{R}^2$ but it is not a s.o.s.\ of elements of $\mathbb{R}[x,y]$.}. While the first condition is NP-hard, the second can be efficiently checked using a SdP, as we are going to show.

Following this approach, the main idea behind our method is to construct linear polynomials $l(\vec{\cal S}_K) \in \mathbb{R}[\vec{\cal S}_K]_1$ satisfying $l(\vec{\cal S}_K) \geq 0$ for all $\vec{\cal S}_K \in {\cal V}$, \textit{i.e.}\ valid Bell inequalities defining half-spaces containing $\mathrm{CH}({\cal V})$. 

Starting from the observation that every polynomial of the form $p + \sum_i f_i p_i$, with $p, p_i \in \mathbb{R}[\vec{\cal S}_K]$, takes the same values when evaluated in ${\cal V}$ (because $f_i(\vec{\cal S}_K) = 0$ for all $\vec{\cal S}_K \in {\cal V}$), we define the ideal $I$ generated by $f_i$ as the set
\begin{equation}
I = \left\lbrace \sum_{i} f_i \, p_i  \;\;\text{s.t.}\;\;  p_i\in\mathbb{R} [\vec{\mathcal{S}_K}] \right\rbrace \subseteq \mathbb{R} [\vec{\mathcal{S}_K}] \;,
\label{eq:def:ideal}
\end{equation}
such that every polynomial in $p + I = \{p + q,\ q \in I\}$ is equivalent when evaluated in $\cal V$. 
Moreover, the ideal $I$ defines the set of equivalence classes $\mathbb{R} [\vec{\mathcal{S}_K}] /I$, where $p,q\in\mathbb{R} [\vec{\mathcal{S}_K}]$ are in the same class if they are equivalent \textit{modulo} $I$, \textit{i.e.}\ $p \equiv q \mod I$, meaning that $p-q\in I$.

To express $l(\vec{\cal S}_K)$ we consider the following \textit{ansatz}:
\begin{equation}
l(\vec{\cal S}_K) = \sum_{i=0}^m g_i(\vec{\cal S}_K) \sigma_i(\vec{\cal S}_K) \mod I		\;,
\label{eq:certificate}
\end{equation}
where $g_0(\vec{\cal S}_K)=1$, and $\sigma_i(\vec{\cal S}_K)$ are s.o.s.\ polynomials \textit{modulo} $I$ (\textit{i.e.}\ there exists a s.o.s.\ polynomial in $\sigma_i({\vec{\cal S}_K}) + I$). For compactness, let us use the shorthand notation $g_i$ and $\sigma_i$.
Note that since all $g_i \geq 0$ in ${\cal V}$ by definition, and s.o.s.\ are non-negative, the form of Eq.~(\ref{eq:certificate}) ensures the non-negativity of $l(\vec{\cal S}_K)$ in $\cal V$, \footnote{The \textit{modulo} $I$ in Eq.~(\ref{eq:certificate}) allows to reduce the degree of $l(\vec{\cal S}_K)$, potentially arriving to a linear polynomial.}. 

Now, given a point $\vec{\cal S}_K^*$, our goal is to prove that $l(\vec{\cal S}_K^*)< 0$ for some $\sigma_i$. If we succeed in this proof, then we have to conclude that $\vec{\cal S}_K^* \notin \mathrm{CH}({\cal V}) \supseteq \mathbb{P}^{\text{S}}_K$, \textit{i.e.}\ that the statistics in $\vec{\cal S}_K^*$ come from nonlocal correlations. 

For computational reasons, we need to bound the maximum degree of the s.o.s.\ decomposition allowed in $\sigma_i + I$. The higher the degree, the larger the family of $l(\vec{\cal S}_K)$ that can be accessed through Eq.~(\ref{eq:certificate}), but the more  computationally expensive to produce such s.o.s.\ representation will be. This naturally yields a hierarchy of outer approximations to $\mathrm{CH}(\cal V)$ by increasing the degree of the s.o.s.\ decomposition of $\sigma_i$. To simplify our exposition, we consider here the special case where all $\sigma_i=\sigma$.

To express all $\sigma$ that are s.o.s.\ of degree $2\mu$, \textit{modulo} $I$, we adopt the following procedure. First, we select (via a Gr\"obner basis \cite{Buchberger}) a linearly independent set of representatives of $\mathbb{R}[\vec{\cal S}_K]/I$, and we order them in the vector $\vec{b} = (1,{\cal S}_0,{\cal S}_1, \dots)^T$. Denoting by $\vec{b}_{\mu}$ the vector of elements of $\vec{b}$ of degree at most $\mu$, we write $\sigma = \sum_{j}s_j^2 \mod I$, where $s_j$ are linear combinations of the elements of $\vec{b}_\mu$; \textit{i.e.}\ $s_j = \vec{b}_\mu^T \vec{a}_j$, with $\vec{a}_j$ real vectors. At this point, by defining the matrix $G = \sum_{j} \vec{a}_j \vec{a}_j^T$, which is positive semi-definite by construction ($G \succeq 0$), and the moment matrix $\Gamma_i = g_i \vec{b}_\mu \vec{b}_{\mu}^T \mod I$, we write
\begin{equation}
g_i \;\sigma =  \Gamma_i \cdot G  \mod I \;, \qquad G \succeq 0 \;.
\label{eq:gsigma}
\end{equation}
Here $ X \cdot Y = \sum_{ab}X_{ab}Y_{ab}$. 

When the elements of $\Gamma_i$ corresponding to $\vec{\cal S}_K$ are replaced by $\vec{\cal S}_K^*$, only some of its entries are constrained. If the remaining free parameters can be tuned to make $\Gamma_i \succeq 0$, Eq.~(\ref{eq:gsigma}) ensures that $g_i \; \sigma  \geq 0$ in $\vec{\cal S}_K^*$ for all $\sigma$ (that are s.o.s.\ of degree $2\mu$, \textit{modulo} $I$). On the other hand, when $\Gamma_i \nsucceq 0$ for any choice of the free parameters, there exists at least one $\sigma$ such that $g_i \; \sigma < 0$ in $\vec{\cal S}_K^*$ \footnote{For a detailed proof, see the proof of Theorem 5.1 in \cite{LasserreBook2}}.

Recall here that our final goal is to prove that there exist a $\sigma$ such that Eq.~(\ref{eq:certificate}) gives $l(\vec{\cal S}_K^*) < 0$. To this end, we write Eq.~(\ref{eq:certificate}) as $l(\vec{\cal S}_K) = \tilde{\Gamma}\cdot\tilde{G} \mod I$, where $\tilde{\Gamma} = \bigoplus_{i=0}^m \Gamma_i$, and similarly for $\tilde{G}$.
As for Eq.~(\ref{eq:gsigma}), we ask whether $\tilde{\Gamma}$ can be made positive semi-definite at the point $\vec{\cal S}_K^*$. To perform this check with a SdP, we first reduce $\tilde{\Gamma}$ \textit{modulo} $I$, and then linearize it as
\begin{equation}
\tilde{\Gamma} = \sum_{j} y_j \tilde{\Gamma}_j  \;,
\end{equation}
where $y_j$ indexes the $j$-th element of $\vec{b}$, and $\tilde{\Gamma}_j$ are constant real matrices embodying the constraints among the entries of $\tilde{\Gamma}$. Now, for the point $\vec{\cal S}_K^*$, we write the SdP
\begin{equation}
\begin{array}{llll}
 \displaystyle\max_{y_j \in \mathbb{R}}& 1 &&\\
 \textrm{s.t.}&\tilde{\Gamma} & \succeq &0\\
 & y_0&=&1\\
 & y_j&=& \;(\vec{\cal S}_K^*)_j
\end{array}
\label{eq:SdPlambda}
\end{equation}
where $y_0$ and the $y_j$ corresponding to $\vec{\cal S}_K^*$ are fixed, while the other $y_j$ are free real parameters that can be varied until the condition $\tilde{\Gamma} \succeq 0$ is fulfilled. 

\begin{figure}
\includegraphics[width=8.6cm]{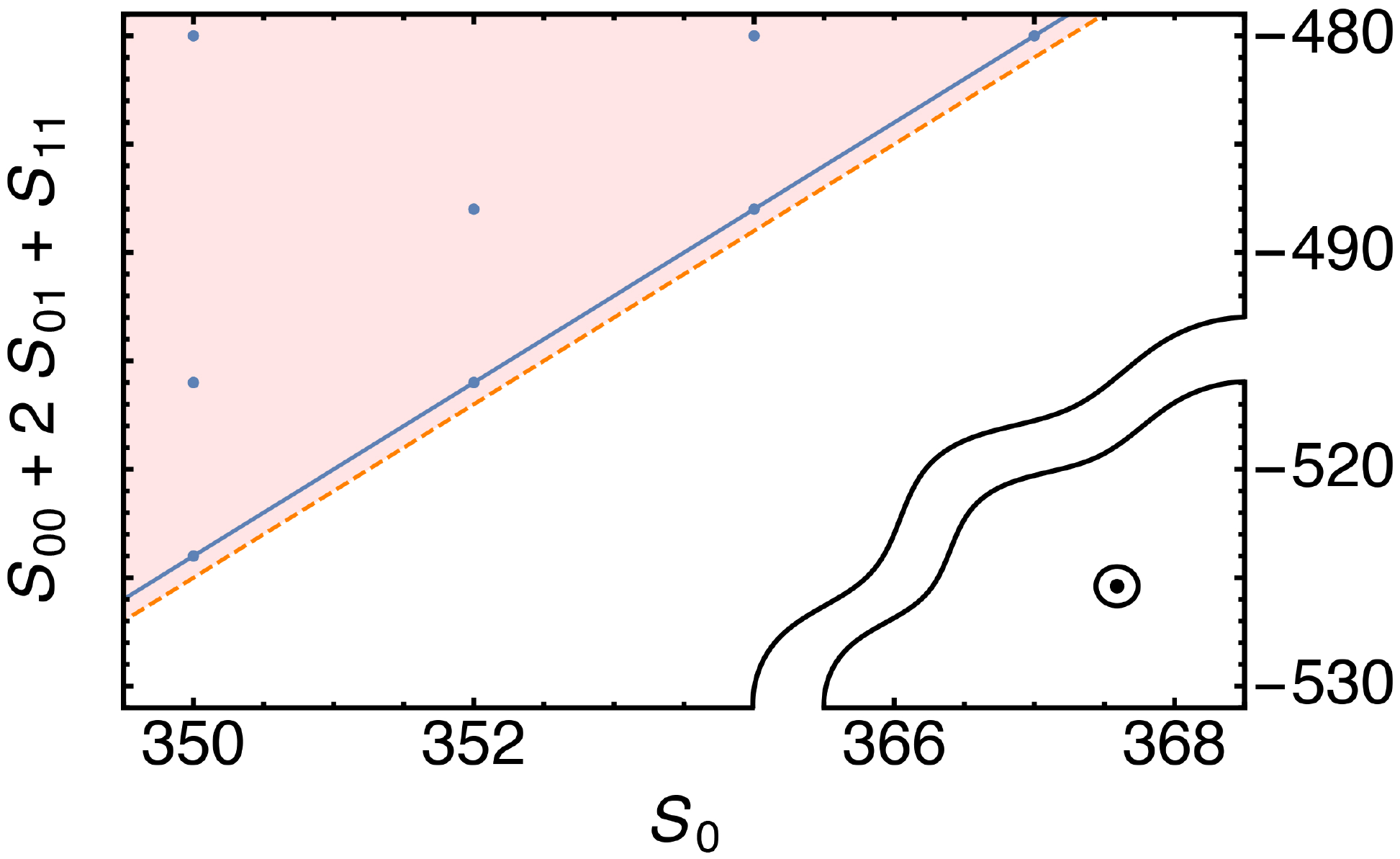}
\caption{\label{fig:2} Plane generated by $\left\lbrace \mathcal{S}_0 , \left( \mathcal{S}_{00} + 2\, \mathcal{S}_{01} + \mathcal{S}_{11}  \right) \right\rbrace $.
Black circled dot, point $\left(367.6, -525.4\right)$ measured experimentally in \cite{SchmiedSCIENCE16} for $N=476$.
Blue points, projected vertices of $\mathbb{P}^{\text{S}}_2$. Blue line, bound given by the Bell inequality $ -2 \mathcal{S}_0 + \left( \mathcal{S}_{00} + 2 \mathcal{S}_{01} + \mathcal{S}_{11}  \right)/2 + 2N \geq 0 $, from \citep{TuraSCIENCE14,SchmiedSCIENCE16}. This inequality is tight, meaning that it is also a facet of the projected polytope.
Pink region, points where SdP (\ref{eq:SdPfeasible}) gives $\lambda\geq 1$.
Orange dashed line, Bell inequality obtained numerically by solving the dual of SdP (\ref{eq:SdPfeasible}).
The distance between the blue and the orange lines is $1.000002$, meaning that the error of our method compared to the tight classical bound scales as $1/N$, and it is imputable mainly to the first relaxation.}
\end{figure}

If SdP (\ref{eq:SdPlambda}) is infeasible, $\tilde{\Gamma} \nsucceq 0$ independently on the free $y_j$, which proves that there exist a $\sigma$ such that $l(\vec{\cal S}_K^*) < 0$. Therefore, infeasiblity of (\ref{eq:SdPlambda}) certifies that $\vec{\cal S}_K^* \notin \mathrm{CH}({\cal V}) \supseteq \mathbb{P}^{\text{S}}_K$, \textit{i.e.}\ its nonlocal nature (see the example and Fig. \ref{fig:2}).

\begin{figure}
\includegraphics[width=8.6cm]{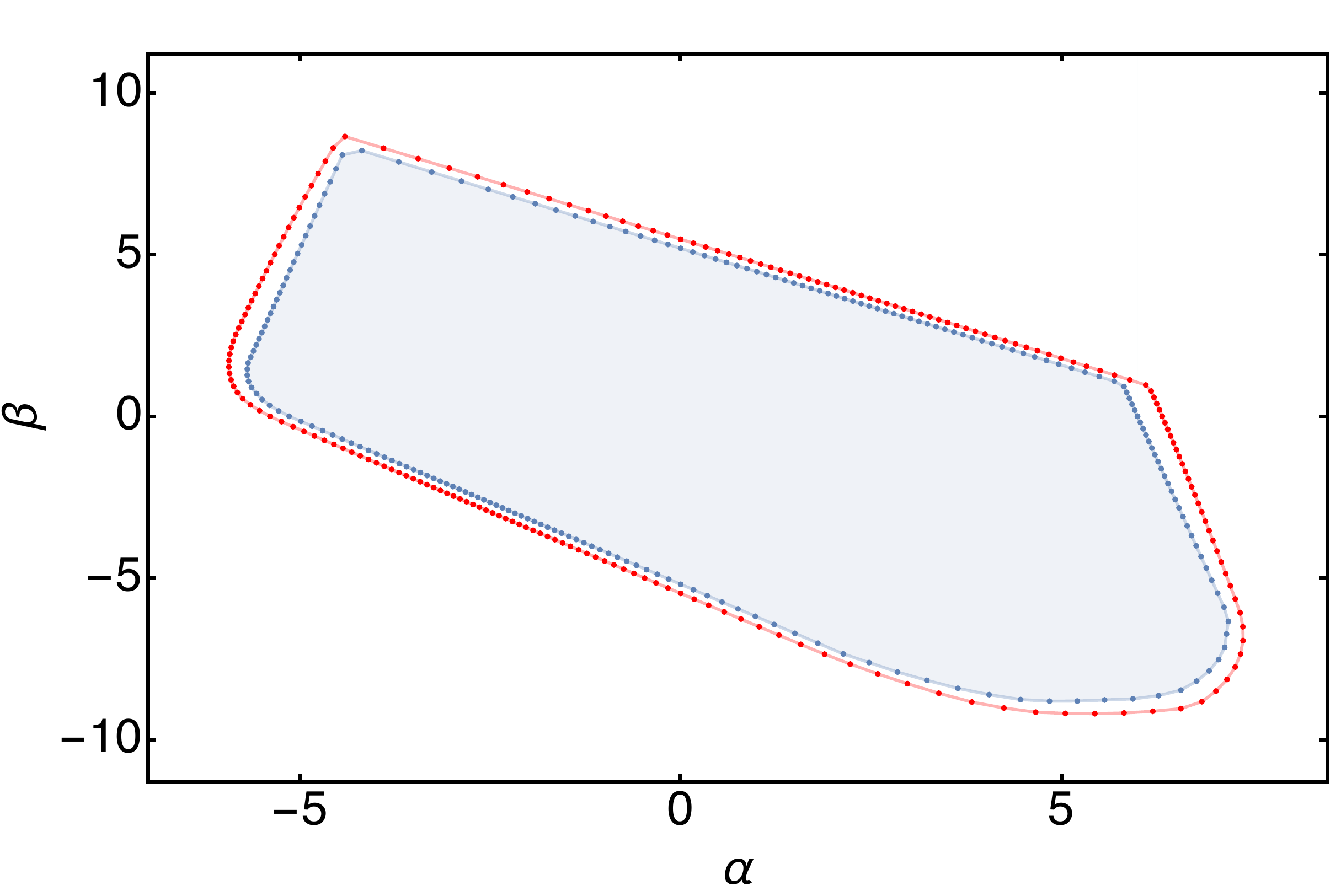}
\caption{\label{fig:1} For $N=10$ and $K=2$, the plane of the symmetric correlations of the form $\alpha \vec{\cal S}_2^{(1)} + \beta \vec{\cal S}_2^{(2)}$, with $\vec{\cal S}_2^{(1)}=(1,-1,0,-1,1)^T/\sqrt{4}$ and $\vec{\cal S}_2^{(2)}=(0,-1,-1,1,0)^T/\sqrt{3}$. In blue, the intersection of $\mathbb{P}^{\text{S}}_2$ with the plane, computed with a linear program. In red, the boundary of the feasible set of SdP (\ref{eq:SdPlambda}) for $\mu = 1$. The gap between the two objects is imputable mainly to the first relaxation, and the small $N$ was chosen also to appreciate its size, which remains of the same order while $\mathbb{P}^{\text{S}}_2$ increases with $N$ (see also Fig. \ref{fig:2})}
\end{figure}

While the output of SdP (\ref{eq:SdPlambda}) is the answer feasible/infeasible, we can also write a SdP to maximize $\lambda$ subject to $y_0=1$ and $y_j =  \;\lambda(\vec{\cal S}_K^*)_j$. The dual formulation of this modified SdP results in the dual variables $\alpha_{j_1 \ldots j_k}$ associated to $y_1 \ldots y_i$, and $\beta_C$ associated to $y_0$, defining a Bell inequality (\ref{eq:BI}) that can be used to certify the nonlocality of $\vec{\cal S}_K^*$ \footnote{See Section 3.1 of \cite{NPA} and, for a detailed derivation, Section 4.2 of \cite{LasserreBook}} (see the example and Fig. \ref{fig:2}). In addition, maximizing $\lambda$ along different directions  $\vec{\cal S}_K^*$ results in the points $\lambda_{\text{max}} \vec{\cal S}_K^*$ that can be used to approximate the boundary of $\mathbb{P}_K^{\text S}$, (see Fig. \ref{fig:1}).

On the other hand, if SdP (\ref{eq:SdPlambda}) is feasible it means that it does not exist a $\sigma$ that is s.o.s.\ of degree $2\mu$, \textit{modulo} $I$, such that $l(\vec{\cal S}_K^*)<0$. In this case, we could access a higher level of our hierarchy by increasing $\mu$, which enlarges the class of $l(\vec{\cal S}_K)$ to be tested \footnote{Note that the matrix $\tilde{\Gamma}$ for the level $\mu$, is a minor of the matrix $\tilde{\Gamma}^\prime$ for the level $\mu^\prime > \mu$. Therefore, if $\tilde{\Gamma}^\prime \succeq 0$ then necessarily $\tilde{\Gamma} \succeq 0$, while the converse is not always true.}.

An additional result in \cite{Gouveia10,GouveiaBOOK12} ensures that, since the variety ${\cal V}$ we want to approximate is compact, our hierarchy converges at least asymptotically to $\mathrm{CH}({\cal V})$. Actually, in all examples we studied, we observed numerically that convergence at $\mu=1$ was already present.

\paragraph{Example.} -- In the spirit of \cite{TuraSCIENCE14,SchmiedSCIENCE16}, we consider $d=K=2$, giving rise to the set of correlators $\vec{\cal S}_2 = ({\cal S}_0, {\cal S}_1, {\cal S}_{00}, {\cal S}_{01}, {\cal S}_{11})\in \mathbb{R}^5$, and $N$ parties. In this scenario, there are four LDS parameterized by $x_i \geq 0$ and satisfying $\sum_{i=1}^4 x_i = N$. By expressing the correlators $\vec{\cal S}_2$ evaluated on a LDS in terms of $\vec{x}$, we obtain \cite{TuraSCIENCE14}
\begin{equation}
 \left(
 \begin{array}{c}
  N\\
  {\cal S}_1\\
  {\cal S}_0\\
  {\cal Z}
 \end{array}
 \right) = 
  \left(
 \begin{array}{c}
  x_1 + x_2 + x_3 + x_4\\
  x_1 + x_2 - x_3 - x_4\\
  x_1 - x_2 + x_3 - x_4\\
  x_1 - x_2 - x_3 + x_4
 \end{array}
 \right) \;,
  \label{eq:Hadamard} 
\end{equation}
\begin{equation}
 \left(
 \begin{array}{c}
  {\cal S}_{00}\\
  {\cal S}_{01}\\
  {\cal S}_{11} 
 \end{array}
 \right) = 
  \left(
 \begin{array}{c}
  {\cal S}_0^2 - N\\
  {\cal S}_0{\cal S}_1 - {\cal Z}\\
  {\cal S}_1^2 - N
 \end{array}
 \right) \;.
  \label{eq:Skl} 
\end{equation}
When $N$ is fixed Eqs.~(\ref{eq:Hadamard}) are three free parameters, while Eqs.~(\ref{eq:Skl}) define the ideal $I$, whose ${d+K \choose d}-m = 2$ generators $\{ f_1(\vec{\cal S}_2), f_2(\vec{\cal S}_2) \} = \{{\cal S}_{00} - {\cal S}_0^2 + N, {\cal S}_{11} - {\cal S}_1^2 + N\}$ form also a Gr\"obner basis for $I$. Inverting Eq.~(\ref{eq:Hadamard}) we obtain four polynomials in $\vec{\cal S}_2$ that allow to express the constraints $x_i = g_i(\vec{\cal S}_2) \geq 0$; \textit{e.g.}\
\begin{equation}
 g_1(\vec{\cal S}_2) = ({\cal S}_0 + {\cal S}_1 + ({\cal S}_{0} {\cal S}_{1} - {\cal S}_{01}) + ({\cal S}_0^2 - {\cal S}_{00}))/4\geq 0 \;. \nonumber
\end{equation}
At the first level of our hierarchy, $\mu = 1$, the vector $\vec{b}_1^T = (1, {\cal S}_0, \ldots, {\cal S}_{11})$ generates the five $6\times 6$ moment matrices $\Gamma_i$. Combined together, the  $\Gamma_i$ give a $30 \times 30$ block-diagonal moment matrix $\tilde{\Gamma}$, in which $N$ appears as a parameter, and thus not affecting its size.

Considering the experimental data presented in \cite{SchmiedSCIENCE16}, we can conclude that the measured statistics $({\cal S}_0^*, {\cal S}_{00}^*+ 2{\cal S}_{01}^*+{\cal S}_{11}^*) = (367.6,-525.4)$ contain Bell correlations because the following SdP gives $\lambda < 1$:
\begin{equation}
\begin{array}{lrll}
 &\displaystyle\max_{y_j \in \mathbb{R}}\  \lambda \;\; & &\\
 & \textrm{s.t.}\quad \tilde{\Gamma} \,\; & \succeq &0 \\
 & y_0 &=& 1 \\
 &(y_1,y_3+2 y_4 + y_5)&=&\lambda({\cal S}_0^*, {\cal S}_{00}^*+ 2{\cal S}_{01}^*+{\cal S}_{11}^*)
\end{array}
\label{eq:SdPfeasible}
\end{equation}

The dual of SdP (\ref{eq:SdPfeasible}) gives as result the dual variables associated to $y_0$, $y_1$ and $y_3+2 y_4 + y_5$, which correspond respectively to the coefficients of the Bell inequality $\beta_C + \alpha_1 {\cal S}_0 + \alpha_2 \left( {\cal S}_{00} + 2 {\cal S}_{01} + {\cal S}_{11}  \right) \geq 0 $, (see Fig. \ref{fig:2}).

\paragraph{Comment on more outcomes} -- It is possible to consider the case where measurements have more outcomes by defining the expectation values as \textit{e.g.}\ $\langle \mathcal{M}_j^{(i)} \rangle^{(a)} = 2 P_i(a\vert j) -1$, where $P_i(a\vert j)$ is the probability that measurement $j$ on party $i$ gives as outcome $a$, and the symmetrized correlators as \textit{e.g.}\ $\mathcal{S}_{j}^{(a)}=\sum_{i=1}^N \langle \mathcal{M}_j^{(i)} \rangle^{(a)}$.

\paragraph{Conclusions.} -- We introduced a method to bound the set of LHVM correlations. Its main advantage, with respect to other techniques, is that there is no scaling with the number of parties, making it particularly suited for the study of nonlocal correlations in many--body systems. Our approach has several applications, some of which were presented here, such as the characterization of experimentally observed correlations or the derivation of new Bell inequalities.
Furthermore, it can be easily generalized to scenarios with more measurements settings and outcomes, potentially enlarging the class of systems, and states, where nonlocal correlations could be experimentally detected.

\paragraph{Acknowledgments.} -- We are grateful to Antonio Ac\'{i}n, Remigiusz Augusiak, Jean--Daniel Bancal, Gemma de las Cuevas, Jo\~{a}o Gouveia and Philipp Treutlein for the useful discussions and comments on the manuscript. MF was supported by the Swiss National Science Foundation. JT was supported by Fundaci\'{o} Privada Cellex through the CELLEX-ICFO-MPQ programme, the Spanish MINECO (SEVERO OCHOA programme for Centres of Excellence in R\&D SEV-20150522 and National Plan FISICATEAMO FIS2016-79508-P), the Generalitat de Catalunya (SGR 874 and the CERCA programme), and EU grants OSYRIS (ERC-2013-AdG Grant No.\ 339106) and QUIC (H2020-FETPROACT2014 No. 641122). This project has received funding from the European Union’s Horizon 2020 research and innovation programme under the Marie Sk{\l}odowska-Curie grant agreement No. 748549.

\bibliography{paperTH}

\end{document}